\documentclass[twocolumn,showpacs,pre]{revtex4}
\usepackage{dcolumn}
\usepackage{graphicx}
\usepackage{graphicx,amssymb,amsmath,revsymb}
\usepackage{natbib}

\begin{document}

\title{Spindles and active vortices in a model of confined filament-motor mixtures}

\author{David A. Head$^{1,2,3}$, W. J. Briels$^{2}$ and Gerhard Gompper$^{1}$}

\affiliation{$^{1}$Theoretical Soft Matter and Biophysics, Institute of Complex Systems, Forschungszentrum J\"ulich, J\"ulich 52425, Germany.\\
$^{2}$Computational Biophysics, University of Twente, 7500 AE Enschede, The Netherlands.\\
$^{3}$School of Computing, Leeds University, Leeds LS2 9JT, United Kingdom.
}

\date{\today}

\begin{abstract}
Robust self-organization of subcellular structures is a key principle governing the dynamics and evolution of cellular life. In {\em fission yeast} cells undergoing division, the mitotic spindle spontaneously emerges from the interaction of microtubules, motor proteins and the confining cell walls, and asters and vortices have been observed to self-assemble in  quasi-two dimensional microtubule-kinesin assays. Their is no clear microscopic picture of the role of the active motors driving this pattern formation, and the relevance of continuum modeling to filament-scale structures remains uncertain. Here we present results of numerical simulations of a discrete filament-motor protein model confined to a pressurised cylindrical box. Stable spindles, nematic configurations, asters and high-density semi-asters spontaneously emerge, the latter pair having also been observed in cytosol confined within emulsion droplets. State diagrams are presented delineating each stationary state as the pressure, motor speed and motor density are varied. We further highlight a parameter regime where vortices form exhibiting collective rotation of all filaments, but have a finite life-time before contracting to a semi-aster. Quantifying the distribution of life-times suggests this contraction is a Poisson process. Equivalent systems with fixed volume exhibit persistent vortices with stochastic switching in the direction of rotation, with switching times obeying similar statistics to  contraction times in pressurised systems. Furthermore, we show that increasing the detachment rate of motors from filament plus-ends can both destroy vortices and turn some asters into vortices. Based on our findings we argue the need for a deeper understanding of the microscopic activities underpinning macroscopic self-organization in active gels and urge further experiments to help bridge these lengths.
\end{abstract}

\pacs{0.0.a}

\maketitle



%
%
\section{Introduction}

Filamentous proteins are prevalent within eukaryotic cells and perform a variety of crucial tasks relating to cellular integrity, locomotion, transport and division~\cite{AlbertsBook,BrayBook}. Such tasks are often {\em active} in that they can only proceed in concert with energy-consuming mechanisms, including directed filament growth and motor protein-generated tension, placing such processes outside the realm of equilibrium thermodynamics~\cite{Mizuno2007}. Self-organisation of motor protein-filament mixtures will be selected for when it robustly reproduces static or dynamic structures beneficial to the cell's viability. An example is the mitotic spindle that forms during division of {\em fission yeast} cells. It has been shown that this bipolar structure, consisting of microtubules emanating from spindle pole bodies towards an overlapping midplane region, exists and functions essentially as normal even in cells with no nucleus-associated microtubule organizing center~\cite{CarazoSalas2006,Daga2006}. The plausible conclusion is that the interaction between filaments and motor proteins in the confined cell geometry controls the location of the pole bodies. For {\em budding yeast} this self-organisation scenario has been reinforced by the evolution of more sophisticated regulatory mechanisms~\cite{Haase2007}, and egg cell extracts from the amphibious genus {\em Xenopus} can generate a well-formed spindle apparatus despite entirely lacking cell walls~\cite{Verde1991,Heald1996,Hentrich2010}. Nonetheless an understanding of the principles underlying self-organization of bioflaments driven by motor proteins in confined spaces is of direct relevance to many organisms~\cite{Terenna2008}.

Given the complexity of real cells it is often advantageous to consider simplified model systems, and this approach has been adopted to investigate the role of confinement in filament-motor mixtures. Experiments on growing microtubules confined to spherical emulsion droplets revealed a droplet-size dependency on the observed structure~\cite{Pinot2009}: Droplets larger than $\approx29\mu m$ in diameter contained {\em asters} with the polar microtubules pointing towards the centre, controlled by the motor protein dynein, whereas smaller droplets were found to contain {\em semi-asters} with the aster's focus near the interface. These findings demonstrate that the degree of confinement can partly determine structure formation, but as motor density and speed were not control variables in these experiments their influence could not be assayed.

A strikingly non-equilibrium property of filament-motor mixtures is their ability to spontaneously generate flows due to their active components, even in the absence of boundary driving forces~\cite{Voituriez2005,Cates2008,Giomi2010}. Assays of microtubule-oligomeric kinesin mixtures in a quasi--two dimensional geometry with flat, parallel confining walls found a dynamic rotating structure denoted a {\em vortex}~\cite{Nedelec1997,Surrey2001}. Accompanying simulations of semiflexible filaments~\cite{Surrey2001} and subsequent hydrodynamic theories~\cite{Kruse2004,Elgeti2011} appeared to reproduce the observed structures. However, as discussed in Ref.~\cite{Head2011}, it is unlikely that the simulations of Surrey {\em et al.}~\cite{Surrey2001} and the theories and simulations of Ref.~\cite{Kruse2004,Elgeti2011} describe the same type of vortex, because the hydrodynamic theories are based on a nematic order-parameter description, while simulations of semi-flexible filaments in Ref.~\cite{Surrey2001} neglect self-avoidance (and thus nematic order). Simulations of self-avoiding filaments strictly in two dimensions showed no evidence of a vortex state~\cite{Head2011}. The microscopic picture underlying vortex formation thus remains unknown. Gliding assays of filaments along motor beds permit quantitative comparison to models~\cite{Hentrich2010,Kraikivski2006} and at high concentration exhibit vortex-like `swirls'~\cite{Schaller2010,Schaller2011}, although in this situation the active forces are unbalanced monopoles, unlike dipoles generated by motors connecting two filaments in the bulk~\cite{Head2010}. Vortex-like motion is often observed in self-propelled systems such as bacterial swimmers~\cite{Czirok1996,Ramaswamy2010,Toner1998}, but with differing microscopic mechanisms.

It is apparent that the combined influence of confinement and activity on structure formation and spontaneous flows in filament-motor mixtures is presently not well understood. Our aim here is to acquire a deeper understanding of this problem in a broad sense, not restricted to any one biological realisation, {\em i.e.} microtubule-dynein or actin-myosin. It is therefore desirable to study model systems in which all parameters can be freely varied. The application of continuum equations, which are coarse-grained over lengths much larger than the filament length~$L$, to structures of only a few $L$ in spatial extent is not guaranteed to be successful. We therefore adopt a discrete numerical model in which motors and filament segments are explicitly represented, and all physical mechanisms that are potentially relevant (steric hinderance, thermal fluctuations {\em etc.}) are incorporated. This model is an extension of one previously employed in two dimensions~\cite{Head2011}, where it was found to produce some signatures of active gels such as super-diffusion and anomalous small wavelength density fluctuations, but not vortices.

We consider arrays of filaments confined to a quasi- two dimensional cylinder, with a height of a few filament diameters which permits filament overlap, and an external pressure at the curved walls. We then systematically vary the motor density, speed and applied pressure. Four steady-state configurations arise within the covered parameter space, including an aster and semi-aster as observed in confined emulsion droplets~\cite{Pinot2009}, and also a spindle-like state that spontaneously emerges from the motor-filament interaction in the confined geometry, possibly reproducing the fission yeast observations~\cite{CarazoSalas2006,Daga2006}. These states are described in Sec.~\ref{s:statics} along with a fourth nematic state that links to known equilibrium phases. We also find a fifth, vortex state associated with a definite rotation of filaments about a fixed center that appears to be always transient. The existence and properties of these vortices are characterised in Sec.~\ref{s:dynamics}. To highlight the important role played by motors at filament plus-ends, we independently vary the detachment rate of motors from plus-ends in Sec.~\ref{s:varykE} and show that vorticity is associated with a critical fraction of plus-ended motors. The observation of vortices in fixed volume systems described in Sec.~\ref{s:fixedVol} confirm that they are driven at least partly by motor motion and not boundary fluctuations, and in Sec.~\ref{s:disc} we discuss possible future directions.

%
%
\section{Model}
\label{s:model}

\begin{figure}
\center{\includegraphics[width=8.5cm]{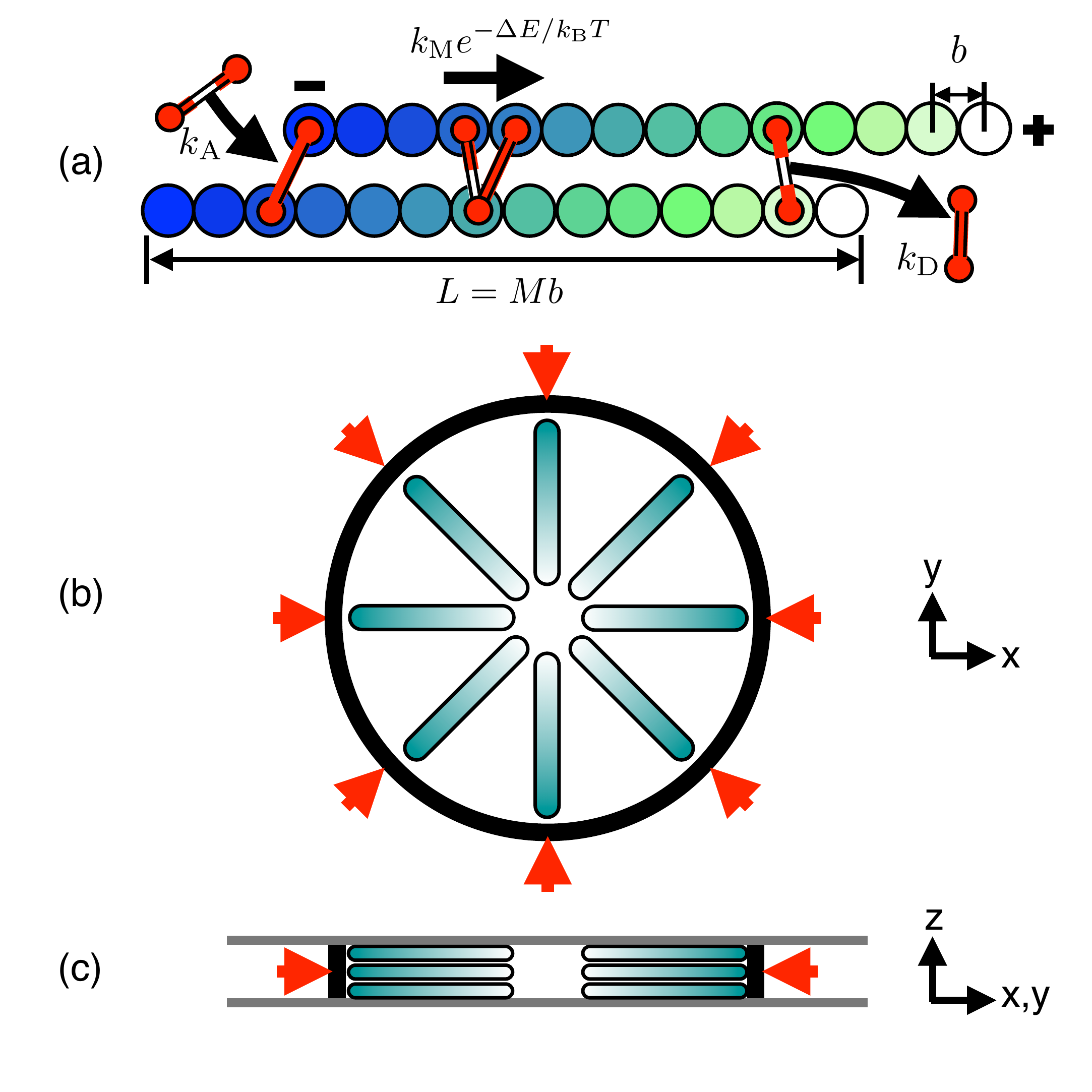}}
\caption{(a)~Summary of key model parameters including the rates of motor attachment $k_{\rm A}$ and detachment $k_{\rm D}$, and the bare stepping rate $k_{\rm M}$. See text for details. (b)~Plan view showing the filaments oriented with their light-shaded $[+]$-ends towards the center. The arrows denote the external pressure acting on the circular elastic wall. (c)~Side view of the same, showing the confining walls perpendicular to the $z$-axis.}
\label{f:modelSchematic}
\end{figure}

We consider a system of $N$ semiflexible polar filaments, which can be connected by motor proteins. Each filament consists of $M=30$ monomers separated by a bond length $b$ with Hookean bond potentials with a spring constant $100\,k_{\rm B}T/b$. Self-avoidance of filaments is introduced by repulsive Lennard-Jones potentials with diameter $\sigma$ and energy parameter $\epsilon=5\,k_{\rm B}T$. A natural choice is $\sigma=b$. Semi-flexibility is described by curvature elasticity with bending rigidity $\kappa=200\,bk_{\rm B}T$ such that the persistence length $\ell_p=\kappa/k_BT$ is $\ell_p=20L/3$ with $L=Mb$ the filament length. Motors are modeled as two-headed Hookean springs with a spring constant $k_{\rm B}T/b^{2}$ and dynamics defined by four rates as shown in Fig.~\ref{f:modelSchematic}(a): (i)~The attachment rate $k_{\rm A}$ for a motor to attach to two monomers within a predefined range; (ii)~the detachment rate $k_{\rm D}$ of each head independently from its filament (detachment of either head results in removal of the whole motor from the system); (iii)~the movement rate $k_{\rm M}$ of each head independently towards the filament's $[+]$-end, and (iv)~the detachment rate $k_{\rm E}$ for motor heads already at a $[+]$ end. The movement rate is attenuated by an exponential factor $e^{-\Delta E/k_{\rm B}T}$ with $\Delta E$ the change in motor spring energy for the trial move. Except where otherwise stated, $k_{\rm E}=k_{\rm D}$ below.

The quasi-2D simulation cell has parallel confining walls normal to the $z$--axis spaced $5b\ll L$ apart, restricting filament orientation to be approximately in the $x$--$y$ plane while permitting a degree of overlap; see Figs.~\ref{f:modelSchematic}(b) and~(c). These walls repel the monomers with the same Lennard-Jones non-bonding potential as for filaments. $N=175$ filaments are placed in a radial aster configuration with all $[+]$-ends pointing towards the center, in three parallel layers with roughly $66$--$68$ filaments per layer (note there is some stochasticity in the initial conditions). This initial condition was chosen to promote the formation of asters and vortices, but does not inhibit other structures as described below. This initial array is surrounded by an elastic wall with both bending and stretching energies, with the elastic coefficients chosen to ensure a circular shape is approximately maintained throughout. The stiffness with respect to changes in wall area is chosen to be sufficiently small such that, for the parameters considered below, wall retraction is countered dominantly by filaments repulsion and not the intrinsic wall elasticity. It does however remove the possibility of wall nodes merging and causing a numerical divergence. Stretching and bending energies were defined similar to above with monomers replaced by wall nodes, with stretching and bending coefficients $k_{\rm B}T/\ell_{0}^{2}$ and $2000\,k_{\rm B}T$ resp., with $\ell_{0}\approx\pi b$ the initial separation between nodes.

The filaments, motors and elastic wall are all updated stochastically. The filaments obey Brownian dynamics~\cite{AllenTildesley} governed by an effective monomer friction coefficient $\gamma$ for hydrodynamically anisotropic slender elements, {\em i.e.} with a 2:1 ratio between implicit solvent drag perpendicular to the filament axis ($=2\gamma$) to the parallel direction \mbox{($=\gamma$)}~\cite{DoiEdwards}. Wall nodes move by Monte Carlo Metropolis moves applied to the $(x,y)$ coordinates of 80 nodes initially equispaced along its contour. The energy for these moves includes the wall elastic and wall-filament interaction energies, and a pressure-volume term $PV$ where $P>0$ is a fixed parameter for each run. To check for convergence with time, various scalar quantities, such as the number of motors per filament, were checked to be constant within noise when plotted against $\log(t)$. In addition, the mean squared rotation $(\Delta\theta)^{2}(t,t+\Delta t)=N^{-1}\sum_{i=1}^{N}|\theta_{i}(t+\Delta t)-\theta_{i}(t)|^{2}$, with $\theta_{i}$ the angle between filament $i$'s centre-of-mass and the nominal centre of the box (more precisely, the mean of all wall nodes) relative to some fixed axis, was checked to no longer to vary with~$t$ to within noise. Stationarity was not achieved for the vortex states, for which alternative measures were employed as described in Sec.~\ref{s:dynamics}.

%
%
\section{Controlled pressure with $k_{\rm E}=k_{\rm D}$}

Results are presented here in terms of the normalised attachment rate $k_{\rm A}/k_{\rm D}$, the normalised motor rate $k_{\rm M}\tau_{b}$ where $\tau_{b}=Lb\gamma/4k_{\rm B}T$ is the approximate time for a filament to freely diffuse one monomer distance; the normalised pressure $P/P_{0}$ with $P_{0}=\varepsilon/b^{3}$ with $\varepsilon=5{\rm k}_{\rm B}{\rm T}$ the Lennard-Jones repulsion energy, and, where relevant, the scaled end-detach rate $k_{\rm E}/k_{\rm D}$. We first describe the observed steady-state configurations before turning to consider the transient vortices.

%
%
\subsection{Stationary states}
\label{s:statics}

\begin{figure}
\centerline{\includegraphics[width=8.5cm]{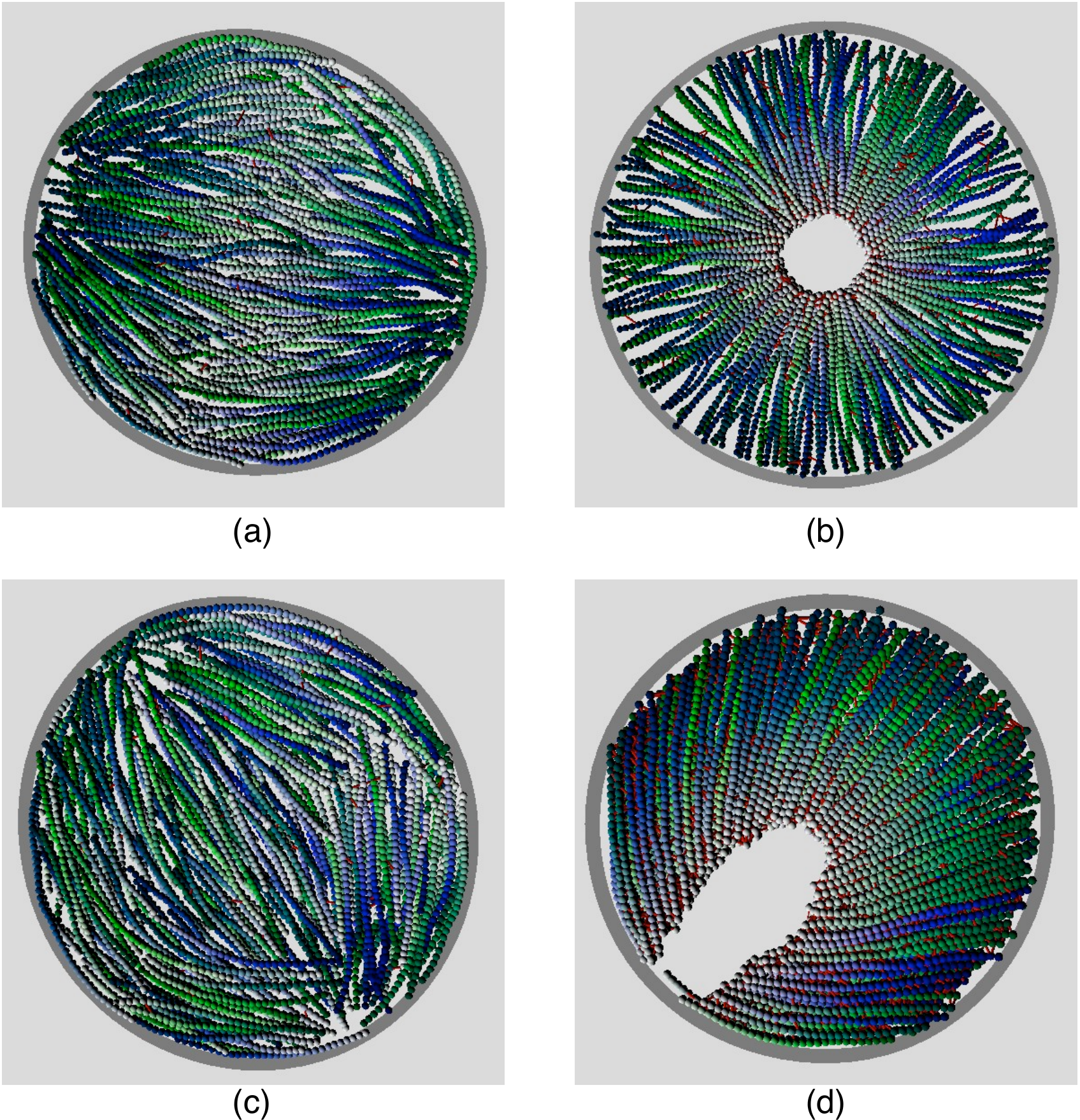}}
\caption{Snapshots of steady-states for a low motor attachment rate $k_{\rm A}/k_{\rm D}=1$ (left) and a higher rate $k_{\rm A}/k_{\rm D}=30$ (right). Conversely, the top line is for fast motors $k_{\rm M}\tau_{b}=3.75\times10^{-2}$ and the bottom line for motors 10 times slower. Filaments are shaded light (dark) towards their plus (minus) ends, respectively. These states are referred to as (a) spindle, (b) aster, (c) nematic and (d) semi-aster. The other parameters are $P/P_{0}=0.03$ and $k_{\rm E}=k_{\rm D}$. Movies of the same parameter values are available from the supplementary information.}
\label{f:snapshots}
\end{figure}

For the parameter space considered, we observe four classes of steady-state configuration as shown in Fig.~\ref{f:snapshots}. For a low density of fast motors, {\em spindles} are observed as in Fig.~\ref{f:snapshots}(a), which crossover to a radially-symmetric {\em aster} as the motor density is increased as show in Fig.~\ref{f:snapshots}(b). For slower motors, we observe a {\em nematic} at low motor densities and {\em semi-asters} at high motor densities, as shown in Figs.~\ref{f:snapshots}(c) and (d) resp. Semi-asters typically arise for higher pressures than asters and are more compressed, consistent with the emulsion experiments of Pinot {\em et al.}~\cite{Pinot2009} and justifying our use of the term. Movies demonstrating the spontaneous emergence of all of these states from the initial conditions are provided in the supplementary information~\cite{SuppInf}, for exactly the same parameters as in Fig.~\ref{f:snapshots}.
  
To quantify to which state a system belongs, each filament's polarity vector is projected onto the $x$--$y$ plane to give a two-dimensional unit vector aligned towards the $[+]$-end. This is averaged over all filaments whose centers of mass have azimuthal angle $\theta$ with respect to the center of the system, giving rise to the mean orientation $\hat{\bf p}(\theta)$. This is then decomposed into angular mode vectors ${\bf a}_{m}$ and ${\bf b}_{m}$,
\begin{equation}
\hat{\bf p}(\theta)=
\frac{1}{2\pi}{\bf a}_{0}+
\frac{1}{\pi}\sum_{m=1}^{\infty}
\left\{
{\bf a}_{m}\cos m\theta+{\bf b}_{m}\sin m\theta
\right\},
\end{equation}
from which can be defined the mode amplitudes $Q_{m}$,
\begin{equation}
Q_{m}=\frac{1}{2\pi^{2}}(a_{m}^{2}+b_{m}^{2}).
\end{equation}
The $Q_{m}$ are invariant under global rotations of the whole box.

To determine the corresponding state, the measured $Q_{\rm m}$ up to $m=3$ are compared to known values for ideal states, and that with the closest Euclidean distance is taken to be the state. The values for pure asters and nematic phases are easy to derive; for an aster $\hat{\bf p}(\theta)=(-\cos\theta,-\sin\theta)$, $(Q_{0},Q_{1},Q_{2},Q_{3})=(0,1,0,0)$, whereas for the nematic state where $\hat{\bf p}(\theta)=(0,0)$, $(Q_{0},Q_{1},Q_{2},Q_{3})=(0,0,0,0)$. Note that while an isotropic state would give the same $Q_{\rm m}$ as for the nematic, such states only arise for $k_{\rm A}$ and $P$ well below the considered ranges. For spindles and semi-aster states there is a degree of choice in how the target $Q_{\rm m}$ are calculated, so we choose simple forms that permit exact evaluation of the $Q_{\rm m}$. For the spindle, $\hat{\bf p}(\theta)=(-\cos\theta,\sin\theta)$ for $\theta\in(-\pi/4,\pi/4)$ or $\theta\in(3\pi/4,5\pi/4)$ and zero otherwise, for which $(Q_{0},Q_{1},Q_{2},Q_{3})=(0,2/\pi^{2}+1/2,0,2/\pi^{2})$. For the semi-aster, $\hat{\bf p}(\theta)=(-\cos[\theta/3],-\sin[\theta/3])$ for $\theta\in(-3\pi/4,3\pi/4)$ and zero otherwise, for which $(Q_{0},Q_{1},Q_{2},Q_{3})=(9/\pi^{2},9/4\pi^{2},333/1715\pi^{2},9/100\pi^{2})$. Variations in these forms have been tested and although the boundaries between states shift slightly, the underlying trends remain the same.

\begin{figure}
\centerline{\includegraphics[width=8.5cm]{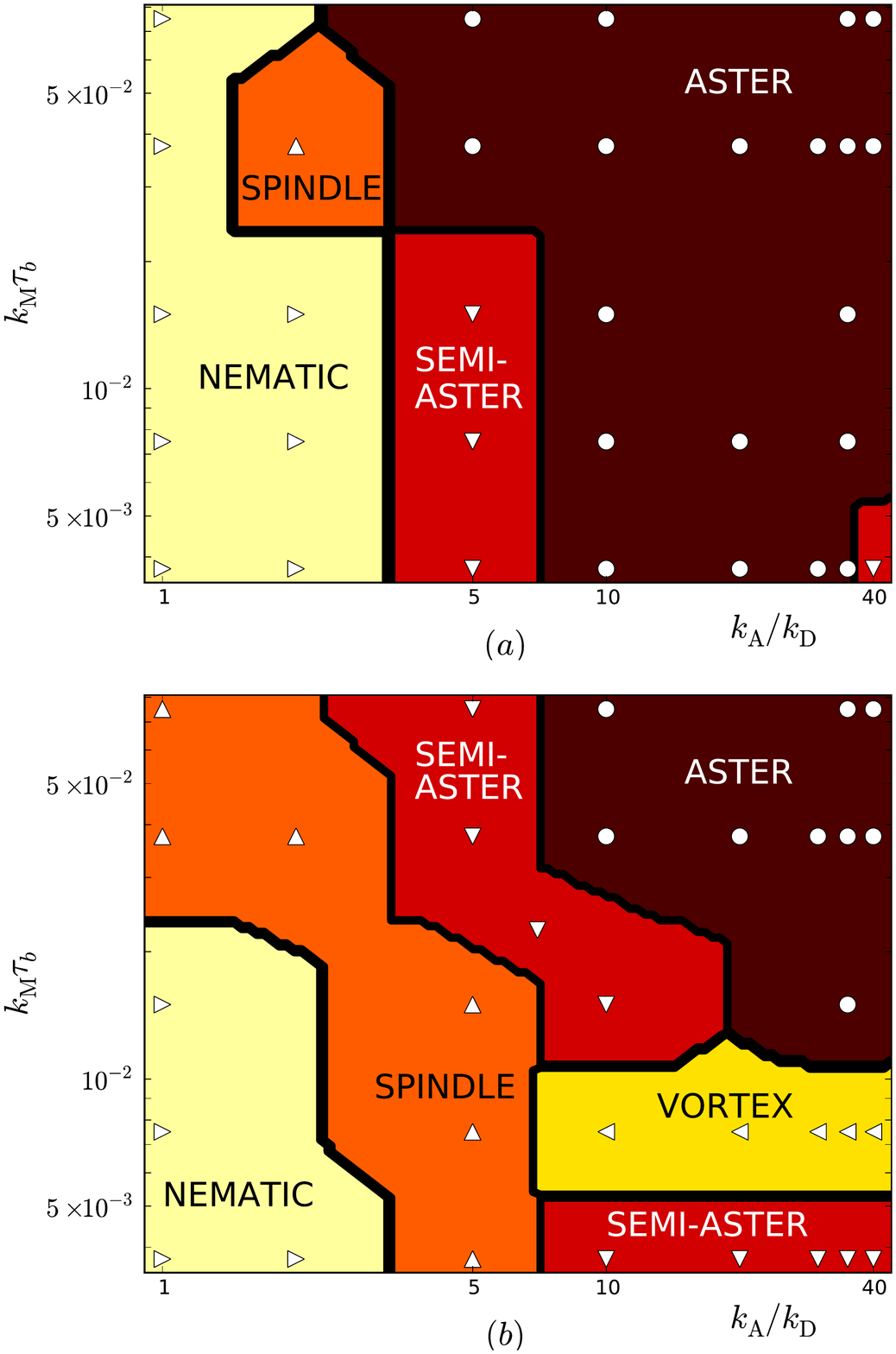}}
\caption{Variation of steady-state with motor density and speed for (a)~$P/P_{0}=0.01$ and (b)~$P/P_{0}=0.024$. Markers denote actual states determined as described in the text and the boundaries are placed midway between points. The vortex region in (b) is a transient configuration that is explained in section~\ref{s:dynamics} and is delineated as those states with a vorticity ${\mathcal V}$ exceeding~0.7. Since it eventually contracts to a semi-aster, the two distinct semi-aster regions in (b) become contiguous in steady-state.}
\label{f:phase}
\end{figure}

\begin{figure}
\centerline{\includegraphics[width=8.5cm]{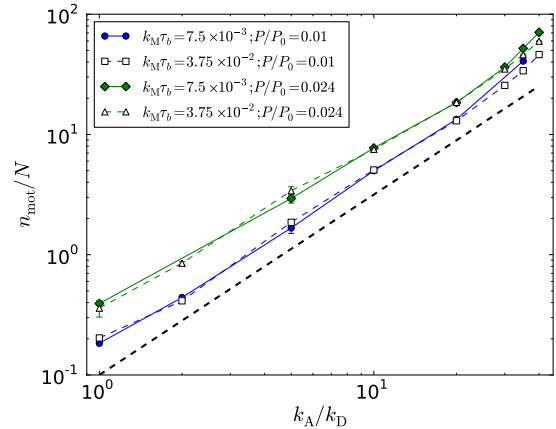}}
\caption{No. of motors per filament {\em versus} attachment rate $k_{\rm A}/k_{\rm D}$ for the motor speeds and external pressure denoted in the legend. The thick black dashed line has a slope of $3/2$. Where data for the required $P$ was not available, $n_{\rm mot}$ was interpolated from runs with $P$ slightly higher and lower than the target value.}
\label{f:motorDensity}
\end{figure}

The occurrence of the four steady-states, plus a fifth `vortex' state to be discussed below, with motor density and speed are presented in Fig.~\ref{f:phase} for two different pressures. The observed configurations correlate with the density and distribution of motors along the filaments. The mean number of motors per filament $n_{\rm mot}/N$ is plotted in Fig.~\ref{f:motorDensity} and shows an expected increase with the attachment rate $k_{\rm A}$ as well as the pressure. The increase with pressure can be understood as due to the closer packing of the filaments, increasing the number of potential attachment points for motors and hence~$n_{\rm mot}$. The approximate scaling $n_{\rm mot}\sim k_{\rm A}^{3/2}$ is faster than the linear relationship measured for constant volume, two-dimensional simulations~\cite{Head2011}, presumably due to similar reasons: As $k_{\rm A}$ increases so does the motor density which, in this constant--pressure ensemble, allows the system to contract, presenting more potential attachment points between monomers and hence further increasing the motor density. A derivation of the value $3/2$ of the exponent is not available so far.

\begin{figure}
\centerline{\includegraphics[width=8.5cm]{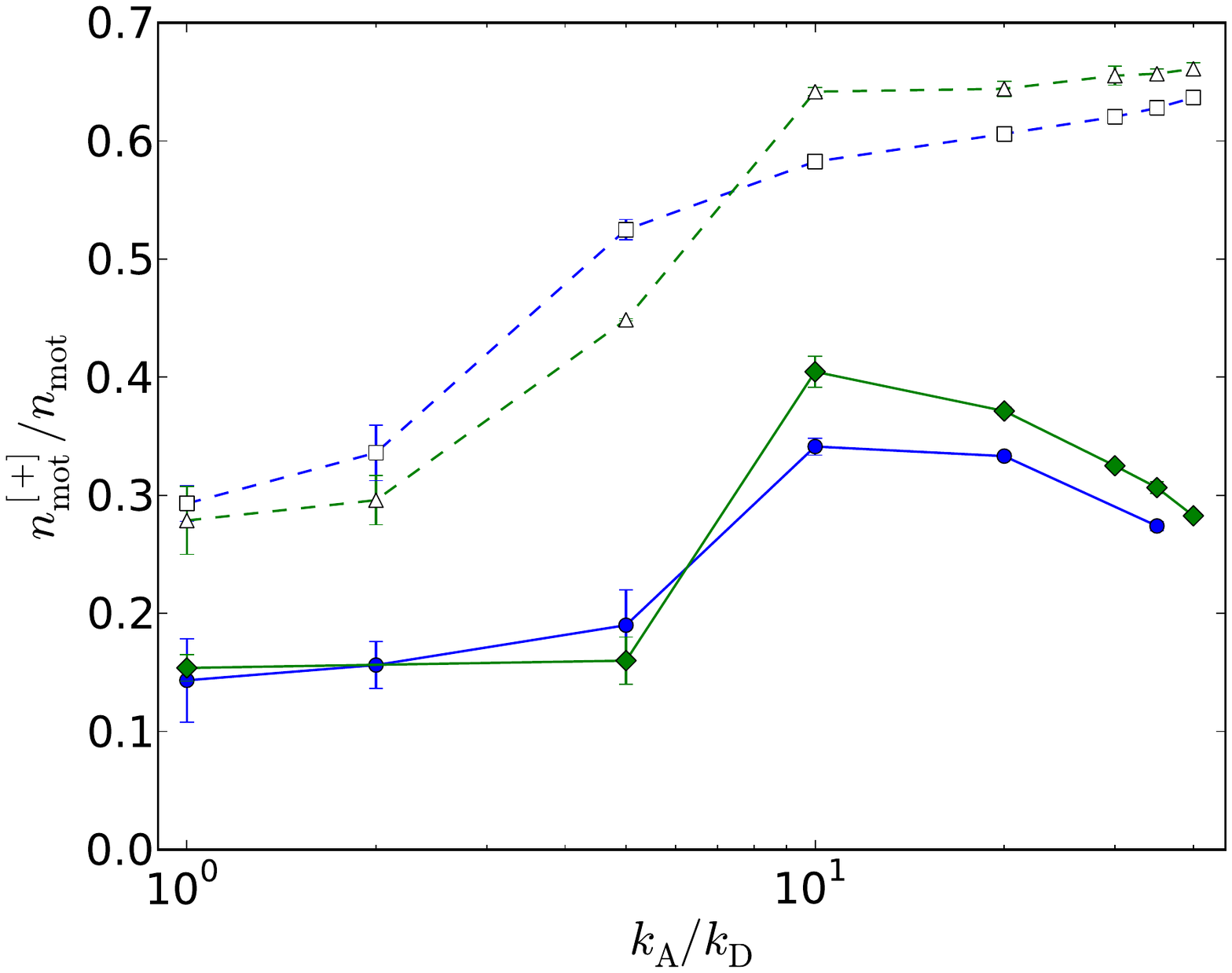}}
\caption{Fraction of motors with at least one head at a filament's $[+]$-end versus $k_{\rm A}/k_{\rm D}$ for the same data as in Fig.~\ref{f:motorDensity}, so the top two lines correspond to fast motors and the bottom to slow motors.}
\label{f:plusEnds}
\end{figure}

Motors move to the $[+]$-end and dwell there until detaching, thus a greater fraction are expected to occupy filament $[+]$-ends, potentially resulting in tight binding mediated {\em via} many motors.
Plotted in Fig.~\ref{f:plusEnds} is the fraction of motors with at least one head at a filament's $[+]$-end, $n^{[+]}_{\rm mot}/n_{\rm mot}$, for the same runs as in Fig.~\ref{f:motorDensity}. By comparing to the configuration plots in Fig.~\ref{f:phase} it is possible to infer signatures of crossovers between states in the inflection points in these curves. For the slower motors, there is an increase in $n^{[+]}_{\rm mot}/n_{\rm mot}$ as the nematic state changes to a state with a greater degree of polar ordering (spindle or semi-aster depending on the pressure). For the faster motors there is a marked increase in $n^{[+]}_{\rm mot}/n_{\rm mot}$ with $k_{\rm A}$, which corresponds to the crossover to the aster state with a high degree of $[+]$--end binding. Further indication of the importance of $[+]$-end binding is presented in section~\ref{s:varykE} where enhanced end-unbinding rates $k_{\rm E}>k_{\rm D}$ are considered.

%
%

%
%

%
%

%
%

%
%
\subsection{Dynamics and vortices}
\label{s:dynamics}

The stationary states described above admit no spontaneous non-equilibrium flows, despite the motor motion generating a positive energy flux: The increase in the stored motor elastic energy due to motor motion and thermal drift of the connected filaments is balanced by the loss due to detachment, with no observed net translocation or rotation of the filaments in steady-state. Collective rotation of all filaments about a fixed center arises for one region of the considered parameter ranges, but appears to be a transient flow that irreversibly contracts to a non-rotating semi-aster configuration. These states are referred to here as {\em vortices} due to their superficial similarity with the rotational modes observed in microtubule-kinesin assays~\cite{Nedelec1997,Surrey2001} and are described in detail in this section. A snapshot is given in Fig.~\ref{f:vortexSnapshot} and movies are provided as part of the supplementary information~\cite{SuppInf}.

\begin{figure}
\centerline{\includegraphics[width=8.5cm]{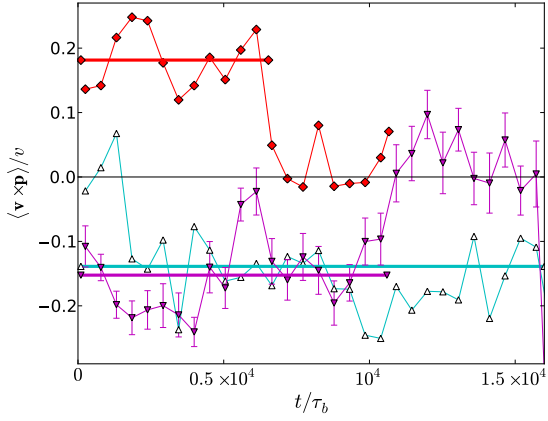}}
\caption{Examples of the vorticity $({\bf v}\times\hat{\bf p})|_{z}$ scaled by a characteristic velocity $v$ for $k_{\rm A}/k_{\rm D}=35$, $k_{\rm M}\tau_{b}=7.5\times10^{-3}$ and $P/P_{0}=0.024$. For clarity error bars are only given for a single run. The thick horizontal line segments denote the mean value up to the time when the vortex contracts to a semi-aster, and are plotted up to this time.}
\label{f:rotation}
\end{figure}

Collective rotation of the whole system can be quantified by the mean angular velocity of filament centre-of-mass vectors ${\bf r}$ relative to the system center, or alternatively by the net transverse velocity of each filament's centre-of-mass relative to its polarity, $\langle({\bf v}\times\hat{\bf p})_{z}\rangle$. Here we employ the latter as it is available for all of our runs, but we have confirmed that it closely tracks the angular velocity in those runs for which both were measured. Examples of ${\bf v}\times\hat{\bf p}$ for 3 independent runs are given in Fig.~\ref{f:rotation}, and show finite rotation of either sign  until the system irreversibly contracts to a semi-aster state and rotation ceases.  This contraction time can be confirmed by visual observation of system states, and can be precisely located by fitting the system radius as a function of time, $R(t)$, to the four-parameter hyperbolic tangent $R(t)=R_{\rm min}+\Delta R\tanh[(t-t^{\rm cont})/\Delta t]$. The mean of $\langle({\bf v}\times\hat{\bf p})_{z}\rangle$ is presented in Fig.~\ref{f:rotation} for each run as a horizontal line segment, that extends from $t=0$ to the contraction time $t^{\rm cont}$ found from this fit. In all cases, $t^{\rm cont}$ coincides with the rapid decay of ${\bf v}\times\hat{\bf p}$ to zero, providing independent confirmation that rotation ceases when the system contracts to a semi-aster.

It is now possible to define a vorticity order parameter for each point in parameter space. For each run~$\alpha$, the mean $\mu_{\alpha}$ and standard deviation $\sigma_{\alpha}$ of $({\bf v}\times\hat{\bf p})|_{z}$ is calculated starting from $t=0$ up to the time that the system contracts. This is regarded as significant if the mean is comparable to or larger than the standard deviation, but since the sign is arbitrary we also take the absolute value to give the {\em vorticity} for a single run,
\begin{equation}
{\mathcal V}_{\alpha}
=\left|\frac{\mu_{\alpha}}{\sigma_{\alpha}}\right|
\geq0.
\label{e:V}
\end{equation}
This is then averaged over all runs with the same parameters to give the mean vorticity ${\mathcal V}=\overline{{\mathcal V}_{\alpha}}$. A given point in parameter space is then regarded as exhibiting a (transient) vortex if ${\mathcal V}$ exceeds some arbitrarily--chosen value of order unity. The corresponding region of parameter space for ${\mathcal V}>0.7$ is plotted in Fig.~\ref{f:phase} and arises for higher densities of motors that are not so fast that they aggregate at $[+]$-ends, which would stabilize an aster relative to a vortex. Independently varying the fraction of $[+]$-ended motors by increasing the end-detachment rate $k_{\rm E}$ supports the existence of a critical fraction for vortex formation, as discussed in Sec.~\ref{s:varykE}.

\begin{figure}
\centerline{\includegraphics[width=8.5cm]{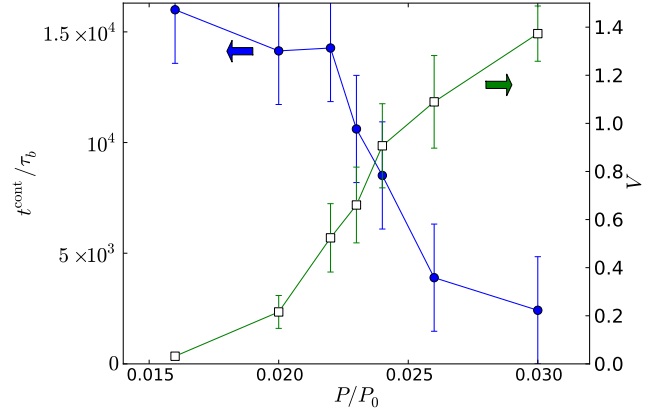}}
\caption{Contraction time to a semi-aster $t^{\rm cont}/\tau_{b}$ (left axis; solid circles) and vorticity ${\mathcal V}$ (right axis; open squares) versus pressure $P/P_{0}$ for $k_{\rm A}/k_{\rm D}=35$ and $k_{\rm M}\tau_{b}=7.5\times10^{-3}$. The contraction time was determined by the fit of the radius to a hyperbolic tangent, or assigned the maximum value of $t^{\rm cont}=1.6\times10^{4}\tau_{b}$ if no contraction had occurred within this time. Each point represents 5 independent runs.}
\label{f:vorticity}
\end{figure}

\begin{figure}
\centerline{\includegraphics[width=8.5cm]{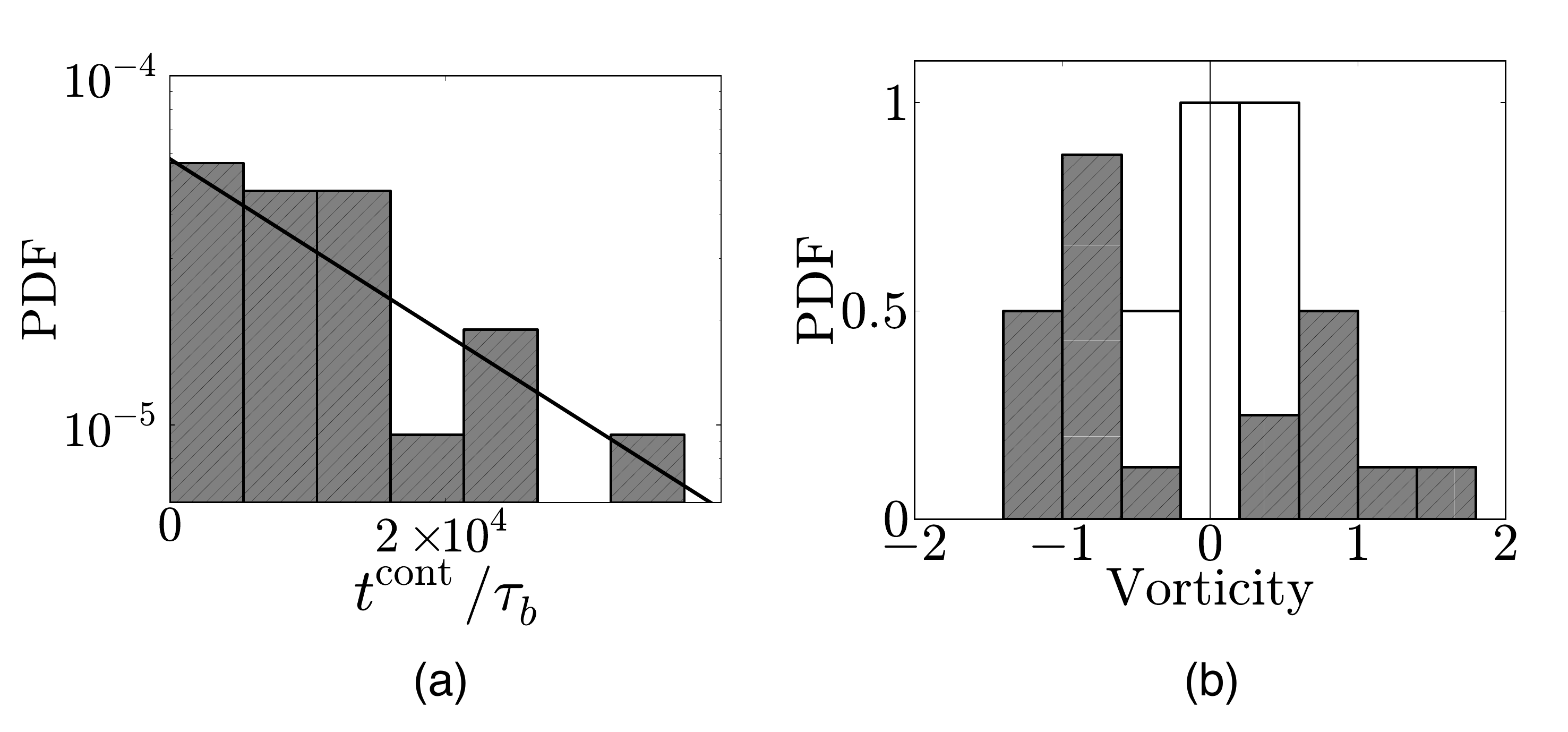}}
\caption{(a)~Probability distribution of contraction times on log-linear axes for $P/P_{0}=0.024$, $k_{\rm A}/k_{\rm D}=35$ and $k_{\rm M}\tau_{b}=7.5\times10^{-3}$. The thick line gives the best fit to an exponential distribution which has a mean $\approx1.7\times10^{4}$, corresponding to $\approx0.75$ full rotations (the longest vortex survived for $\approx2.6$ rotations). Data corresponds to 20 independent runs. (b)~Normalised probability histogram of signed vorticity for $P/P_{0}=0.020$ (white bars in the background; 5 runs) and $P/P_{0}=0.024$ (shaded bars in the foreground; 20 runs).}
\label{f:vortHist}
\end{figure}

The reciprocal relationship between vorticity and contraction time is clearly evident when both quantities are plotted together; see Fig.~\ref{f:vorticity}. Stronger vortices have a shorter lifetime than weaker vortices. The distribution of contraction times is presented in Fig.~\ref{f:vortHist}(a) for a single pressure, and although noisy it appears to be roughly consistent with an exponential form suggesting that contraction is a Poisson process. To give some measure of the goodness-of-fit, the Anderson-Darling statistic for an exponential distribution with an unknown mean has a significance level of $P\approx0.2$, whereas that for a normal distribution of unknown mean and variance has a significance level of $P\approx0.025$~\cite{Spurrier1984,Stephens1986}. Thus we cannot discard the hypothesis that the underlying distribution is exponential. Attaining even this noisy data consumed considerable computing resources and we were unable to repeat this procedure for other parameter values.

Assuming the true distribution is exponential, this would suggest that contraction is triggered by spontaneous fluctuations that occur at a constant rate in time. From observation of movies of filament arrangements, a likely candidate is the transient void formation  frequently observed near the outer wall, where nearby filaments are attached purely by motors at their $[+]$-ends and not along their length. Such voids, when large enough, lead to a `hinge'-like mechanism in which the void expands and one section of the polarity field inverts, leading to the semi-aster.

The onset of vorticity is also evident in the histogram of the {\em signed} vorticity, {\em i.e.} the ${\mathcal V}_{\alpha}$ before taking the modulus in eqn.~(\ref{e:V}), which can be positive or negative depending on the direction of rotation. For low pressures with ${\mathcal V}\ll1$ this distribution is unimodal around the origin, but becomes bimodal when vorticity is more evident as demonstrated in Fig.~\ref{f:vortHist}(b). Of the 20 runs presented here, 12 rotated in one direction and 8 in the other, which has a significance interval of $P\approx0.5$ as determined from a Binomial test with equal probabilities for both directions. This is to be expected given our use of stochastic initial condition that does not predispose the system to any preferred rotational direction.

\begin{figure}
\centerline{\includegraphics[width=8.5cm]{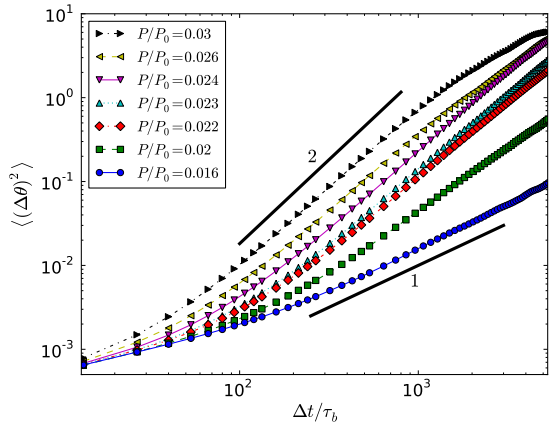}}
\caption{Mean-squared changes in angle $\langle(\Delta\theta)^{2}\rangle$ versus lag time $\Delta t$ for the pressures given in the legend, $k_{\rm A}/k_{\rm D}=35$ and $k_{\rm M}\tau_{b}=7.5\times10^{-3}$. The short thick line segments have the slope given.}
\label{f:angleMSD}
\end{figure}

Independent confirmation of vorticity can be inferred from the mean-squared angular deviation $\langle(\Delta\theta)^{2}\rangle$ already defined in Sec.~\ref{s:model}. This is plotted in Fig.~\ref{f:angleMSD} for the same parameters as above as a function of the lag time $\Delta t$, averaged over all waiting times $t$ up until~$t^{\rm cont}$. There is a crossover from linear behavior $(\Delta\theta)^{2}\sim\Delta t$ for low pressures with low vorticity, to a more rapid scaling $(\Delta\theta)^{2}\sim(\Delta t)^{2}$ for pressures well into the vortex regime. Since this quantity is the angular analogue of the mean squared displacement for translation degrees of freedom, these two limits can be regarded as {\em diffusive} and {\em ballistic}, respectively. Microscopically the diffusive limit corresponds to fluctuations with no net drift, whereas the ballistic limit arises when all filaments are rotating around the system with a constant angular velocity in the same direction. Therefore the vortex state should correlate with ballistic motion, and comparison of Figs.~\ref{f:vorticity} and~\ref{f:angleMSD} confirms this. This figure also demonstrates that the integrated angular rotation of the vortex before contraction is typically larger than $\pi/2$, much larger than the diffusive drift $\approx\pi/10$ over the same time frame.

%
%
\begin{figure}
\centerline{\includegraphics[width=6cm]{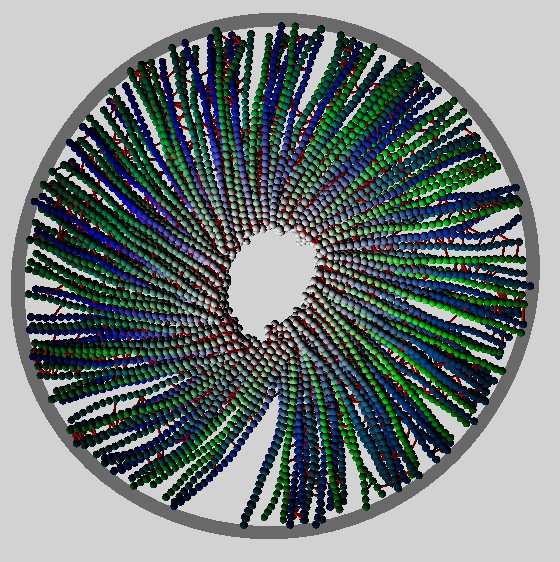}}
\caption{Snapshot of a vortex rotating in the anti-clockwise direction as presented, for parameters $P/P_{0}=0.024$, $k_{\rm A}/k_{\rm D}=35$ and $k_{\rm M}\tau_{b}=7.5\times0^{-3}$ taken at a time $t/\tau_{b}\approx5.3\times10^{3}$. The color code is the same as in Fig.~\ref{f:snapshots}.}
\label{f:vortexSnapshot}
\end{figure}

%
%

%
%

%
%

%
%

%
%
\section{Enhanced detachment from ends}
\label{s:varykE}

\begin{figure}
\centerline{\includegraphics[width=8.5cm]{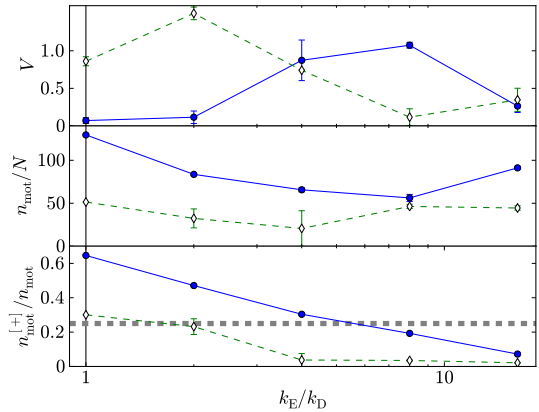}}
\caption{Variation of vorticity, motor density and fraction of $[+]$-end motors with $k_{\rm E}$ for $k_{\rm A}/k_{\rm D}=35$ and $k_{\rm M}\tau_{b}=7.5\times10^{-3}$ (solid lines, filled squares) and $k_{\rm A}/k_{\rm D}=60$ and $k_{\rm M}\tau_{b}=37.5\times10^{-3}$ (dashed lines, open diamonds). $P/P_{0}=0.024$ in both cases. The thick dashed line in the lower plot corresponds to 25\%. Quantities were measured just prior to contraction, or in steady-state if there was no contraction or it happened too rapidly to discern.}
\label{f:varykE}
\end{figure}

In the simulations that accompanied the microtubule experiments, it was claimed that the residence time at the microtubule $[+]$-ends played a crucial role in determining the vortex stability, with an enhanced end-detachment rate  required to form vortices~\cite{Surrey2001}. By contrast, for our model it is the {\em fraction} of motors at filament $[+]$-ends that determines vortex stability relative to an aster. The vorticity, motor density and fraction of motors at $[+]$-ends are plotted in Fig.~\ref{f:varykE} against end--detachment rates $k_{\rm E}\geq k_{\rm D}$ for two sets of $k_{\rm A}$, $k_{\rm M}$ and $P$. It is clear that increasing $k_{\rm E}$ can both destroy a vortex that existed when $k_{\rm E}=k_{\rm D}$, and create a vortex when $k_{\rm E}=k_{\rm D}$ gave an aster. In order of increasing $k_{\rm E}$, the sequence aster $\rightarrow$ vortex $\rightarrow$ semi-aster (where any vortex is either absent or too short lived to be discerned) is typically observed, although we do not claim this sequence is followed by all points in parameter space. There is a slight increase in motor density in the semi-aster state as evident from the figure, resulting from an increase in potential attachment points due to the increased density.

Thus residence time at $[+]$-ends, which is $\propto k_{\rm E}^{-1}$, is {\em not} the determining factor with regards vortex stability here. Rather, vortices coincide with around 25\% of motors at  $[+]$-ends as highlighted in the figure. There is no critical dependency on the motor density, although we speculate that below some minimum value spindles or nematic states would be observed instead. The critical fraction 25\% will likely depend on parameters that were not varied in this work, such as filament length $M$ and the motor spring stiffness. A systematic survey of these parameters, or of $k_{\rm E}\neq k_{\rm D}$ for all $k_{\rm A}$, $k_{\rm M}$ and $P$, is however beyond the scope of this work.

%
%
\section{Controlled volume}
\label{s:fixedVol}

One message from the previous sections is that the observed steady-state is predominately determined by the density of motors and the fraction at $[+]$-ends. It may appear that the primary role of motor motion, which would be the source of any non-equilibrium effects in this model, is merely to select the fraction of $[+]$-ended motors, faster motors giving a higher fraction. It might even be speculated that even the transient vortex state is driven, not by motor motion, but rather as a protracted buckling event powered by the pressurised walls.

\begin{figure}
\centerline{\includegraphics[width=8.5cm]{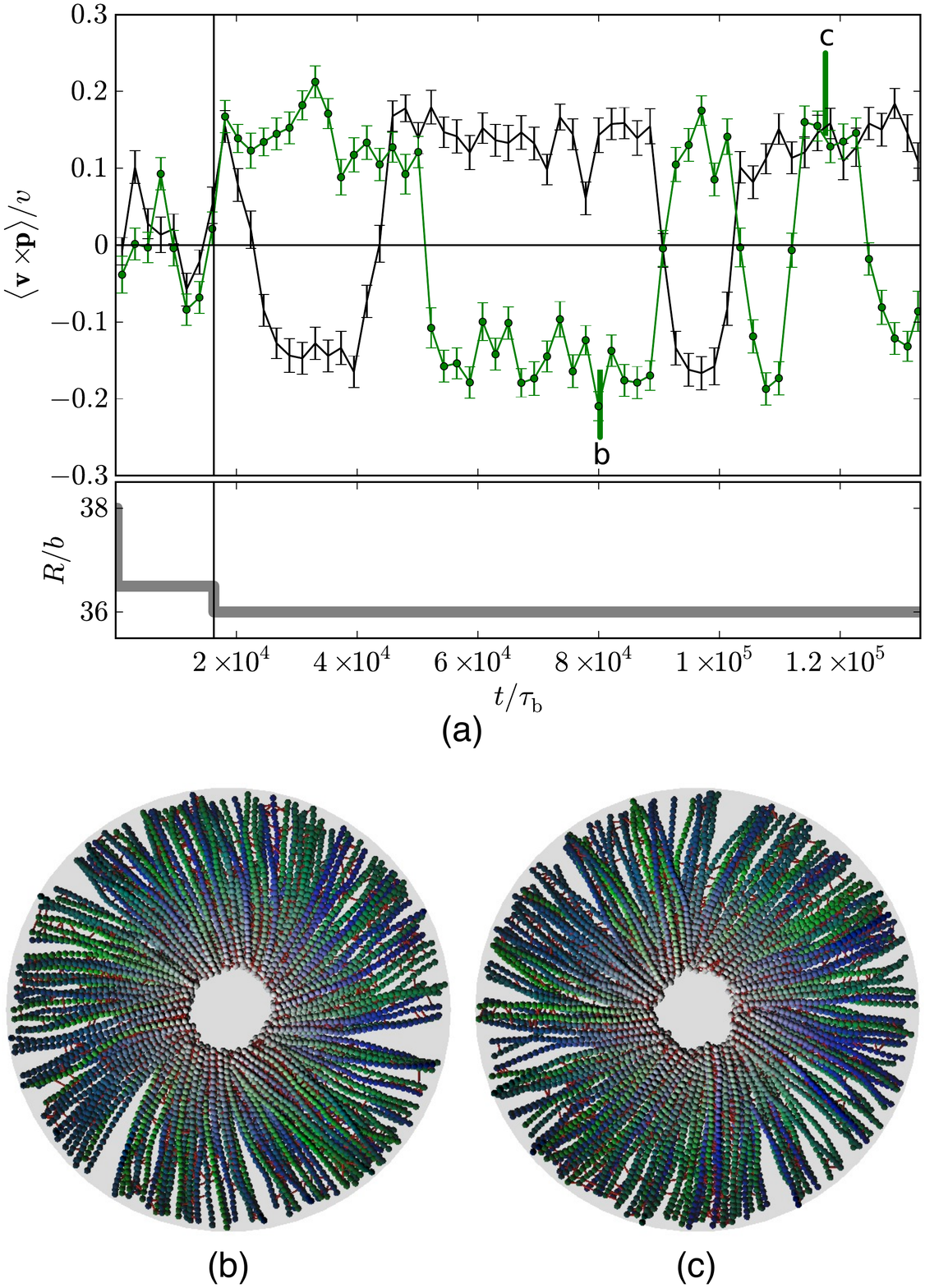}}
\caption{(a)~Filament rotation for two independent runs at fixed volume. The imposed radius $R$ is given in the lower panel. Snapshots for the run corresponding to the solid black line at points (b) and (c) are given in the lower figures. A short time at the initial radius $R/b=38$ was required to avoid numerical instabilities.}
\label{f:fixedVol}
\end{figure}

It is straightforward to show that motor motion can drive vortex motion, however. Plotted in Fig.~\ref{f:fixedVol} is the rotational velocity $\langle {\bf v}\times\hat{\bf p}\rangle/v$ for two independent runs in a box with fixed radius, where $v$ is a characteristic filament velocity. The system is initially in an aster configuration, but when the radius is suddenly reduced by $b/2$ at a time $t/\tau_{b}\approx1.6\times10^{4}$, the system switches to a rotating vortex state that appears to be long-lived; the total time window in this figure is an order of magnitude longer than the longest vortex described in Sec.~\ref{s:dynamics} (which has the same parameters). Since there is no energy input from the walls, the only possible cause for this rotation is the motor motion. Thus the pressure ensemble is important to let the system adjust its density to the vortex state; however, the same pressure also destabilizes the vortex state, because it favors further contraction into the semi-aster.
       
Although the magnitude of the rotational velocity remains fixed (note that the characteristic velocity $v$ is the same for both runs and constant in time), the direction aperiodically reverses as evident in the changes of sign in the figure. The statistics of time intervals between direction switching suggest that the underlying mechanism may be the same as for the contraction to the semi-aster state in the constant pressure case. Specifically, the mean switching time $\Delta t^{\rm switch}/\tau_{b}=21.6\times10^{4}\pm5.8\times10^{4}$ is consistent with the mean contraction time $\approx 1.7\times10^{4}$ measured earlier, and again is consistent with an exponential distribution (significance level $P\approx0.2$ from $n=7$ values using the Anderson-Darling test~\cite{Spurrier1984,Stephens1986}). This suggests that the same spontaneous fluctuation that permits contraction under constant pressure instead promotes rotational reversal under constant volume.

%
%
\section{Discussion}
\label{s:disc}

It has been demonstrated that the vortices described here involve the collective rotation of the filaments about a fixed centre. This was predicted by the nematodynamics theory of Kruse {\em et al.}~\cite{Kruse2004}, but contrasts with the simulations of intersecting filaments~\cite{Surrey2001} for which movies appear to show no actual filament rotation, rather the motors run along a {\em static} vortex configuration of the growing filaments. It is possible that our inclusion of excluded volume interactions, which are necessary to generate the nematic elasticity required by the theory but were absent in these earlier simulations, may explain this discrepancy. Those simulations also employed growing filaments, whereas here as in the continuum modeling these lengths were fixed, suggesting a further potential source of discrepency. Unlike the continuum theories, however, our vortices are only one filament in radius so there is no radial gradient in polarity field, making direct comparison problematic. We conclude that the challenging task remains to demonstrate a definitive link between macroscopic vortices and microscopic filament-motor interactions. An important aspect could be the system size of the microscopic models, because a minimum size much larger than currently accessible by simulations may be needed to see vortices in a bulk system. Further experiments might elucidate the underlying mechanism. For instance, fluorescently tagging a small fraction of microtubules would allow individual filament rotation (if any) to be visualised.

The variation of steady-state structure with motor speed and density shown in Fig.~\ref{f:phase} can in large part be understood as due to a non-uniform density of motors along the filament, in particular the fraction of motors at a $[+]$-end which dwell there before detaching. Varying the end-detachment rate $k_{\rm E}$ confirms that a strong binding at $[+]$-ends can stabilise an aster relative to a vortex or semi-aster. The motor speed $k_{\rm M}$ plays a role in selecting the distribution of motors along the filament, but also contributes to the rotation of vortices as inferred from the fixed volume system in Sec.~\ref{s:fixedVol}. Therefore we claim that the observed vortex is a genuine non-equilibrium state powered at least partially by the unidirectional motion of energy-consuming motor heads along the filaments, although at constant pressure they appear to be transient. It is not clear if varying some other parameters may produce stable vortices.

Systematically quantifying the role of all of the model parameters is clearly challenging for such a high-dimensional parameter space, and here we have adopted the pragmatic approach of holding most parameters fixed while varying those deemed most likely to be critical. Eventually the impact of all parameters on structure and dynamics will need to be quantified if a broad description of active gels is to be attained. Here, we highlight two parameters likely to reveal novel or interesting behaviour. First, the filament length $L=Mb$ was fixed at $M=30$ monomers throughout, whereas extensive simulations with $M=25$ revealed similar steady-state diagrams as Fig.~\ref{f:phase} but {\em no vortices}~\cite{unpub}. Increasing the aspect ratio therefore seems to enhance vorticity, and it would be interesting to quantify this effect. Secondly, the elastic parameters for the wall were set to maintain a roughly circular shape, as in the emulsion experiments of Pinot {\em et al.}~\cite{Pinot2009}. However, in those experiments flexible vesicles were also considered that produced a richer array of observed structures, and this effect could be easily investigated within our model by lowering the bending stiffness of the wall.


%
%
\section*{Acknowledgements}

Financial support of this project by the European Network of Excellence ``SoftComp'' through a joint postdoctoral fellowship for DAH is gratefully acknowledged.


\end{document}